\documentstyle[epsf,aps]{revtex}

\def \bi{\bibitem}

\def\e{{\rm e}}

\def\d{{\rm d}}
 \def\(({\left(}
 \def\)){\right)}
\def\bi{\bibitem}

\def \ov{\over}

\def \b{\beta}

\def \d{{\rm d}}
\def \beq{\begin{equation}}
\def \eeq{\end{equation}}
\def \ln{{\rm ln}}
\def \ov{\over}

\def \ol{\overline}

\def \b{\beta}

\def \ln{{\rm ln}}

\def \ab2{\alpha\beta^2}

 \newcommand {\be} {\begin{equation}}
\newcommand {\bea} {\begin{eqnarray} \nonumber }
\newcommand {\ee} {\end{equation}}
\newcommand {\eea} {\end{eqnarray}}
 \newcommand {\eps} {\epsilon}

\newcommand {\del} {\delta}

\input{epsf}
\begin{document}
\title{Glass transition
and effective potential
in the hypernetted chain approximation}

\author{Miguel Cardenas (*),  Silvio Franz(**) and Giorgio Parisi(***)\\
}
\address{
(*) Scuola Normale di Pisa\\
Piazza de Cavalieri 7,
56126 Pisa, Italy\\
(**) Abdus Salam International Center for Theoretical Physics\\
Strada Costiera 11,
P.O. Box 563,
34100 Trieste (Italy)\\
(***) Universit\`a di Roma ``La Sapienza''\\
Piazzale A. Moro 2, 00185 Rome (Italy)\\
e-mail: {\it cardenas@SABSNS.sns.it, 
franz@ictp.trieste.it, giorgio.parisi@roma1.infn.it}
}
\date{November 1997}
\maketitle

\begin{abstract}
We study the glassy transition for simple liquids in the 
hypernetted chain (HNC) approximation by means of an effective 
potential recently introduced. 
Integrating the HNC equations for hard spheres,  
we find a transition scenario 
analogous to that of the long range disordered systems with 
``one step replica symmetry breaking''. Our result agree 
qualitatively with Monte Carlo simulations of three dimensional 
hard spheres. 
\end{abstract}

\section*{}


The HNC approximation is one of the most widely used approach to describe the density-density 
correlation function $g(x)$ for liquids at equilibrium \cite{macdo}.  It consists in a 
self-consistent integral equation that can be derived by a partial resummation of the Meyer 
expansion, and corresponds to the variational equation for a suitable free-energy functional 
\cite{61,MEPA}.  The simple HNC approach does not by itself allow to detect freezing \cite{vetro}.  
The simple inspection of the pair correlation function certainly does not allow to do so, being 
qualitatively similar in the liquid and in the glass.  Freezing, although present, can be hidden if 
one concentrates on simple equilibrium quantities \cite{remi}. 

In has been recently stressed in \cite{MEPA} that the freezing transition can be detected combining 
the HNC approximation with the replica method by studying the correlation functions among different 
replicas of the same system in presence of a potential which couples them.  At low temperature (or at 
high density) one finds a self-consistent solution where different replicas remains correlated also 
in the limit of zero coupling.  This phenomenon correspond to freezing and it goes under the 
technical name of replica symmetry breaking.

In this letter we pursue this idea of studying the glass transition in the HCN approximation.  We 
are not concerned about the behavior in the glassy phase.  Our aim it to use an effective potential 
recently introduced by two of us \cite{I,letter}, to study the glass transition of HNC hard spheres 
in 3D.  We compare the results with Monte Carlo simulations of real hard spheres.  The conceptual 
advantage of this approach is that all the subtle points of the usual approach related to replica
symmetry breaking are not needed in order to expose the transition.

The effective potential is constructed as follows, for a system described 
by the coordinates of all the particles $x=(x_1,...,x_N)$ and with 
potential energy  $H(x)=\sum_{i<j}^{1,N}\phi(x_i-x_j)$.
 
Let us consider a reference configuration $y$ chosen with probability $\exp(-\b' H(y))/Z(\b')$, 
where $\b'=1/T'$ is some arbitrary inverse temperature.  Let us define a distance among 
configuration as $d(x,y)=1-q(x,y)$, with the ``overlap'' $q(x,y)$ defined as $q(x,y)={1\over N}
\sum_{i,j}^{1,N}w(|x_i-y_j|)$.  $w$ is an attractive potential that in this letter we choose as 
$w(r)=\theta(r_0-r).$ To very different configurations it corresponds large distance and small 
overlap, to similar configurations small distance and large overlap.  We define a constrained 
Boltzmann-Gibbs measure at temperature $T$ as
\be 
\mu (x|y) = {1\ov Z(\b,q,y)} \ \e^{-\b H(x)} \ \delta (q(x,y)-q)
\label{meas}
\ee
where $Z(\b,q,y)$ is the integral over $x$ of the numerator. 
This conditional measure allows to probe regions of the configuration 
space having vanishingly small probability, and as we will see, it 
will help us to reveal the  glassy structure hidden in the 
simple equilibrium approach. 

Introducing a Lagrange multiplier conjugated to $q$ to enforce the 
delta function and integrating over it by saddle point, one see that 
the free energy associated to (\ref{meas}),
$V(q)=-T \log Z(\b,q,y),$
can be computed as the Legendre transform of 
$F(\eps)=-T \log Z(\b,\eps,y)$ with
$Z(\b,\eps,y)= \int \d x \  \e^{-\b( H(x)-\eps q(x,y))}.$
If the coupling $\eps$ is positive there is an attraction 
to the reference configuration $y$. Of special interest will be 
the cases $\eps\to 0^+$, while $q$ will go to a non trivial value. 
The free-energy $F(\eps)$ and the potential $V(q)$ should be 
self-averaging with respect to the distribution of $y$, and be therefore 
just functions of their argument and the temperatures $\b$ and $\b'$. 
From now on, in this letter we will limit ourselves to 
the case $\b=\b'$ which will be enough to detect freezing in the system. 
It is conceptually important however to consider the more general case 
if one would like to describe a system which, after
crossing the freezing temperature, remains confined in the 
vicinity of the configuration where it was last able to thermalize.

In order to compute $F(\eps)$ in any physical system we need to average 
$Z(\b,\eps,y)$ over the distribution of $y$.  This can be done in a convenient way using the replica 
method, where one writes $\ol{\log Z}=\lim_{r\to 0}{\ol{Z^r}-1\ov r}$, and compute the limit from an 
analytic continuation from integer $r$.  In principle the replica method can be avoided but it is 
quite useful to make all the computation quite straightforward.  Explicitly:
\be 
\ol{Z^r}=\int \d x_0 \d x_1...\d x_r \ \e^{-\b \sum_{a=0}^r H(x_a)+
\b\eps \sum_{a=1}^r q(x_0,x_a)}
\ee
we have written $x_0=y$.
The problem is reduced to that of an equilibrium mixture 
of $r+1$ species (with $r\to 0$), 
and is formally similar to the one developed by 
Given, Stell and collaborators
to study liquids in random matrices \cite{given}. 
The use of the formalism is however different. In \cite{given} the 
replica method was used to deal with the quenched disorder represented 
by the medium, 
while for us the potential is a tool to 
probe regions of configuration space  of small Boltzmann probability
and we do not have quenched disorder. 
The HNC equation can be derived from the following free-energy 
functional \cite{61,MEPA}
\be
-2\b F(\eps) = \int \d^d x \sum_{a,b=0}^r  \rho^2 g_{ab}(x)
\left[ \log g_{ab}(x) -1 +\b \phi(x)\delta_{ab}\right] 
+2\b\eps\sum_{a=1}^r \rho^2 g_{0a}(x)w(x)+{\rm Tr}\ {\bf L}(\rho h)
\label{fhnc}
\ee
with 
${\bf L}(u)=u-u^2/2-\log(1+u)$, $h_{ab}=g_{ab}-1$, the trace 
of ${\bf L}$ is intended both on replica indices and in the 
operator sense in space. 
The equations, and the relative value of the free energy are obtained 
extremizing (\ref{fhnc}) over all the replica correlation functions
$g_{ab}$'s, and extracting the terms of order $r$. 
\begin{figure}
\begin{center}
\epsfxsize=350pt
\epsffile{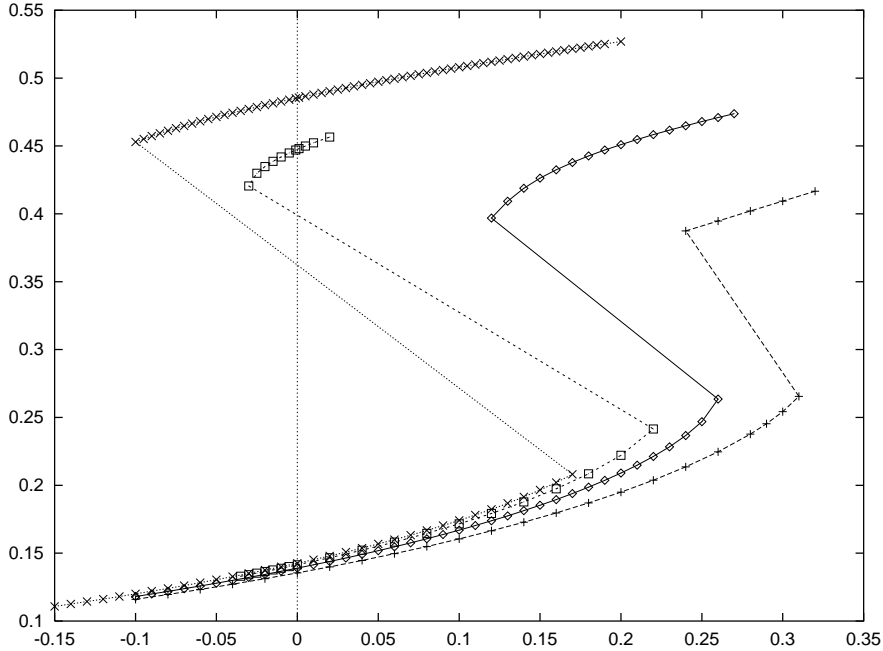} 
\end{center}
   \caption[0]{\protect\label{q-eps}  
The behavior of $q$ as a function of $\eps$ for 
HNC hard spheres for $\rho=1.14,1.17,1.19,1.20$. 
For high enough density $q$ is a multivalued function of $\eps$. 
We have shown only a portion of the curve in the region where it is multivalued.  For graphical 
transparency in this and the next figure we have joined with a line the branches corresponding to 
the same density.  }
\end{figure}

In order to continue analytically $F$ we use the ansatz $g_{ab}=g_{00}$ for $a=b=0$, $g_{ab}=g_{10}$ 
for $a=0$ or $b=0$ and $a\ne b$ and $g_{ab}=g^*_{ab}$ for both $a$ and $b$ different from 0.  In this 
paper we will only consider the replica symmetric ansatz 
$g_{ab}^*=g_{11}\del_{ab}+g_{12}(1-\del_{ab})$.  We warn the reader that this ansatz should  give 
the correct value of the potential $V(q)$ for high and low values of $q$ in the liquid phase, replica 
symmetry breaking effects are to be expected in an intermediate regime \cite{bfp} even in the 
liquid phase.  The physical meaning of the various elements of $g_{ab}$ within this ansatz is 
immediate.  The element $g_{00}$ represents the pair correlation function of the free system; as 
such the equation determining it decouples from the other components in the limit $r\to0$ and 
coincides with the usual HNC equation for a single component system.  In turn, 
$g_{11}$ represents the pair correlation function of the coupled system.  $g_{10}$ is the pair 
correlation among the quenched configuration and the annealed one, while $g_{12}$ represents the 
correlation between two systems coupled with the same quenched system.  This last is the analogous 
of the Edwards-Anderson order parameter in disordered systems \cite{MPV}, and represents the long 
time limit of the time dependent autocorrelation function at equilibrium \cite{I}.
\begin{figure}
\begin{center}
\epsfxsize=350pt
\epsffile{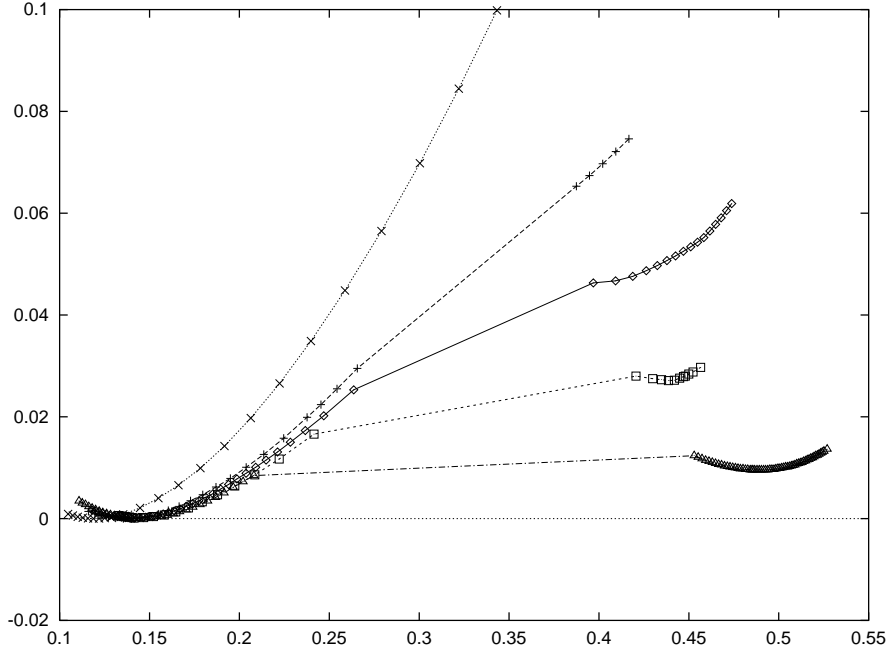} 
\end{center}
   \caption[0]{\protect\label{pot}  
The effective potential  for 
HNC hard spheres. From to to bottom $\rho=1.0,1.14,1.17,1.19,1.20$. For low 
density, high up in the liquid phase the potential is convex. 
In the glass phase two minima are present.} 
\end{figure}
In order to study $q$ as a function of $\eps$ we have solved the HNC equations in 3D with the hard 
sphere potential $\phi(r)=\infty $ for $r<1$, $\phi(r)=0 $ for $r>1$ for various values of the 
density (using a space resolution equal to .01 and a large distance cutoff equal to 10).  The 
density $\rho$ is the control parameter of the freezing transition in this problem where there is no 
temperature.  We have rescaled $\b \eps
\to
\eps$, and chosen $r_0=0.3$ in the definition of the overlap.  We have reconstructed the curves 
$q(\eps)$ and $V(q)$ following the solution of the HNC equations starting from high and low values 
of $\eps$ and respectively decreasing or increasing it slowly.  We see in figure
\ref{q-eps} and \ref{pot} that for low enough density $q$ is a 
single valued function and the potential is convex, with the minimum 
corresponding to $g_{01}(x)=1$ for all $x$. This is a fair description 
of the  liquid phase.
Above a critical density $\rho^*\approx 1.14$ 
the potential loose convexity, and a coupling
can induce a transition between a high and a low $q$ phase \cite{letter}. 
For $\rho=\rho_c\approx 1.17$
a second minimum at high $q$ appears, deepening and deepening as the
 density is increased. The presence of this second minimum shows that 
above $\rho_c$ the system with  is in the glassy phase. If, by means of 
a large $\eps$ we prepare the system in the vicinity of $y$ and we then let 
$\eps\to 0$, the system remains confined. It should be noticed that while
the shape of $V(q)$ depends on the particular definition of the overlap,
the properties of the minima of the potential do not, as they correspond to 
vanishing coupling. 

\begin{figure}
\begin{center}
\epsfxsize=350pt
\epsffile{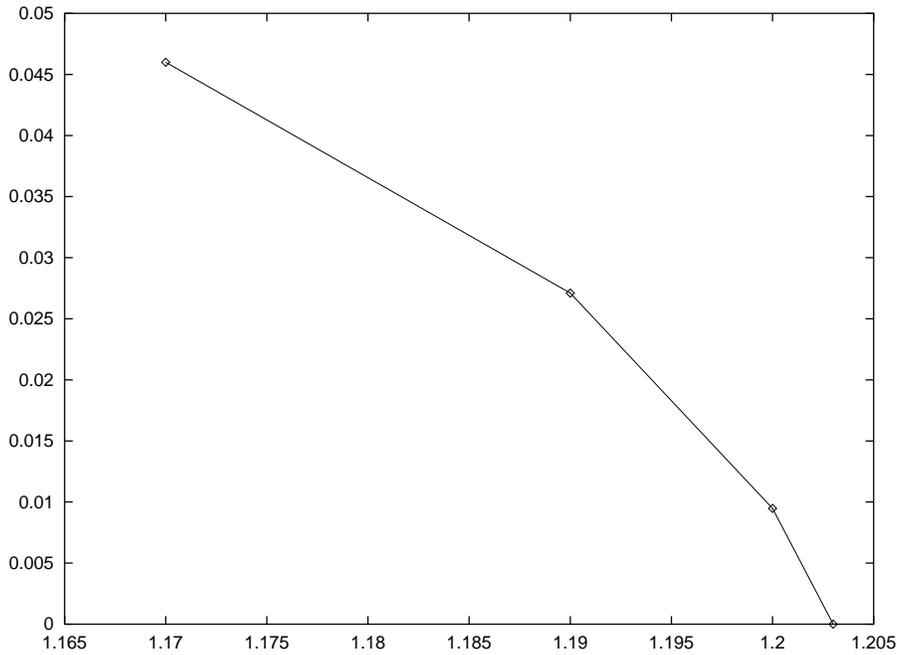} 
\end{center}
   \caption[0]{\protect\label{sigma}  
The configurational entropy $\Sigma$ as a function of $\rho$.} 
\end{figure}

\begin{figure}
\begin{center}
\epsfxsize=350pt
\epsffile{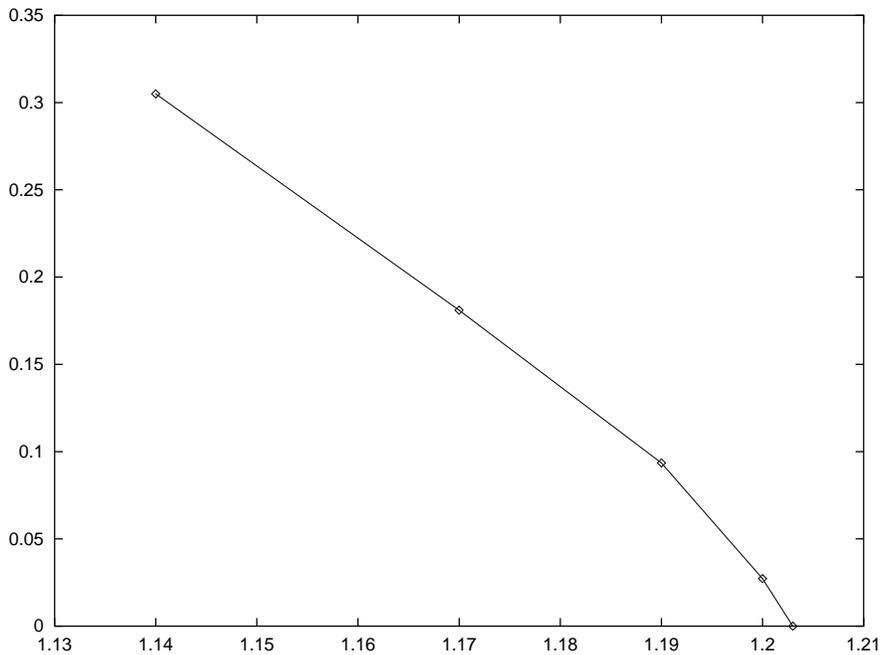} 
\end{center}
   \caption[0]{\protect\label{ph_dia}  
Phase diagram in the plane $\eps-\rho$. A first order transition 
line terminating in a critical point
separates a low $q$ from a high $q$ phase.}
\end{figure}

As it has been discussed in 
\cite{I,letter}, the two minima structure is associated to a Gibbs-Di Marzio 
glass transition scenario. At $\rho_c$ the ergodicity is broken and 
an extensively large number of metastable states
${\cal N}=\e^{N\Sigma}$ contributes to the 
partition function. The relative height of the two minima is 
exactly equal to the ``configurational entropy''  $\Sigma$, and  as we can see in fig. \ref{sigma},
as $\rho$ is increased $\Sigma$ decreases, until it 
vanishes at $\rho=\rho_s\approx 1.203$.  (The values of $\rho_{c}$ and $\rho_{c}$ are compatible 
with those found in \cite{MEPA}, indeed the potential method reproduces the results of replica 
symmetry braking approach for the static and dynamic critical densities).  The shape of the potential 
is the characteristic one of a system undergoing a first order phase transition.  We can use Maxwell 
construction to locate the transition line in the plane $\eps-\rho$, which is shown in fig.
\ref{ph_dia}.  The computation as it stands is not consistent for $\rho>\rho_{s}$: it gives a negative 
configurational entropy in that region. To describe consistently the behavior there
 the replica symmetry breaking formalism of \cite{MEPA} is needed. 
Although we did not try to compare quantitatively the values of 
the freezing density we get with the one previously  found
 in numerical simulations
\cite{HANSEN},
we have performed our own Monte Carlo simulations to test in a qualitative way the prediction of a 
first order transition in presence of a coupling with a fixed configuration.

\begin{figure}
\begin{center}
\epsfxsize=350pt
\epsffile{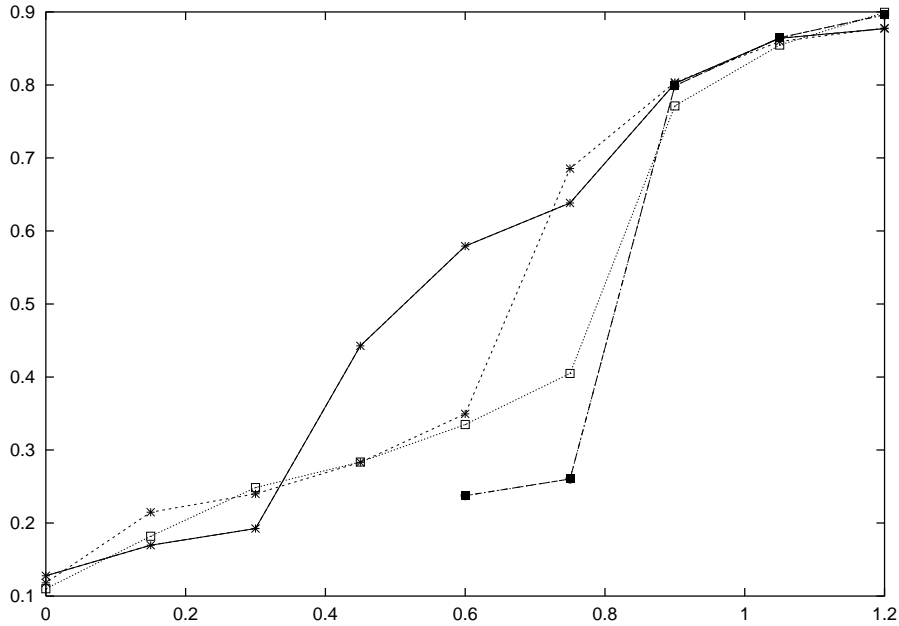} 
\end{center}
   \caption[0]{\protect\label{A}  
The behavior of $q$ as a function of $\eps$ for a system of 258 
particles and $\rho=1.04$. The different curves correspond 
different thermalization times $2^k$ for each value of $\eps$.
>From top
to bottom  $k=17,19,21,23$. For larger thermalization times the system
seem to develop a first order jump in $q$. 
 } \end{figure}

To generate configurations at fixed density we start with $N$ particles 
of zero radius in a three dimensional
 box with periodic boundary conditions, and 
we let the radii grow until two particle do get in contact. At that point we 
make a Monte Carlo sweep and iterate the procedure until the desired
 density is reached. The volume and the radius ($r$) are
 at the point rescaled 
in order to have $r=1$. We thermalize then the system for 4000 Monte
Carlo sweeps and use the configuration $y$ reached as ``external field'' 
for our coupled replicas experiments. The relatively 
short thermalization is chosen in order to avoid crystallization. 
Having generated the configuration $y$ we let evolve a 
coupled system $x$. For various densities, we start the evolution
from the configuration $y$ with an high
value of $\eps$ and decrease the value of $\eps$ in units of $\delta\eps$, 
making $2^k$ Monte Carlo iterations for each value of $\eps$. 
In figure \ref{A} we plot $q$ as a function of 
$\eps$ for different values of $k$. 
We see that, as it should be expected for a system undergoing a first order 
phase transition, the curves are smooth for low $k$ and tend to develop 
a discontinuity for large $k$.  We have presented here results for $\rho=1.04$.  Other 
simulations(which we do not display here) show lower density the discontinuity occurs at higher 
$\eps$, while it is pushed toward smaller $\eps$ for higher density.  From the quantitative point of 
view there is about a 20\% agreement on the value of the density at which a transition is present 
in the $\eps$-$q$ plane; however the qualitative prediction of a first order transition at vales of
$\eps$ of order 1 is clearly satisfied.

\begin{figure}
\begin{center}
\epsfxsize=350pt
\epsffile{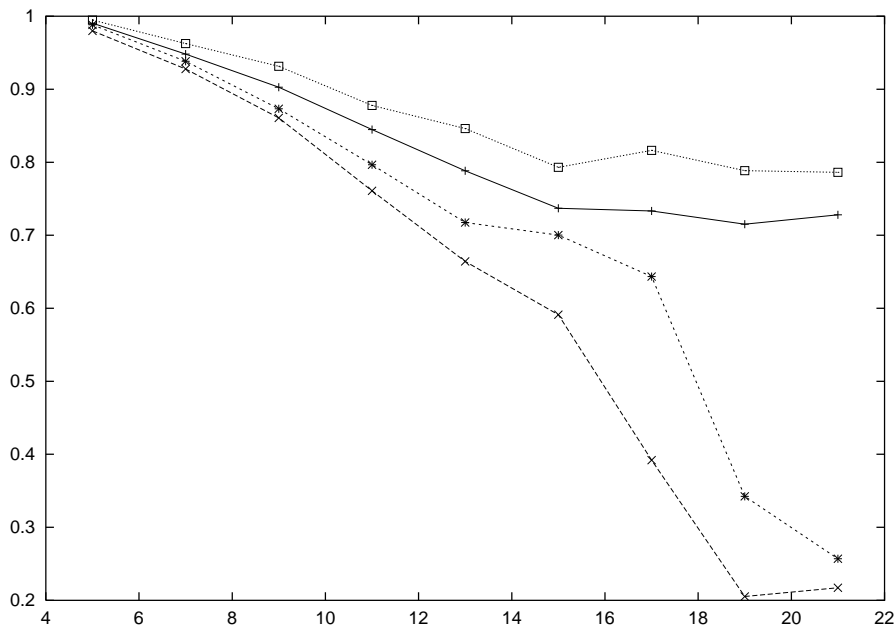} 
\end{center}
\caption[0]{\protect\label{E} The overlap as a function of time in a logarithmic scale (the 
horizontal axis is $\ln_{2}(t)$ starting from $y$ at time 0 and evolving for fixed $\eps$.  In this 
figure $\rho=1.04$ and from top to bottom $\eps=0.8,0.7,0.6,0.5$, the number of particles is 1024.  
} \end{figure}

A different numerical experiment is presented in figure \ref{E}. 
Here we let the system evolve at fixed $\eps$ starting at time zero 
from $x=y$ and we plot 
the overlap as a function of time. 
Again we observe a behavior compatible with a discontinuity of 
$q$ as a function of $\eps$. 

We see that the HNC approach predicts a glass transition scenario very close to the one found in 
systems with ``one step replica symmetry breaking'' with a non convex \cite{I} effective potential 
and Gibbs-Di Marzio entropy crisis \cite{kirtir}.  The HNC is in this respect a genuine mean-field 
theory, giving infinite life meta-stable states.  In real systems metastable states have finite life 
time and the potential has to be a convex function of $q$ for any densities.  As it has been 
discussed many times in the application of the theory to real systems \cite{parisi} the picture 
should be corrected to take that into account.  The density $\rho_c$, representing the point where 
the relaxation time diverges in mean-field, becomes a crossover value where the dynamics is 
dominated by barrier jumping processes \cite{vetro}.  Elucidation of the dynamical processes 
responsible for restoration of ergodicity beyond mean-field is one of the currently open issue in 
glass physics.

\section*{Acknowledgments} S.F. thanks the ``Dipartimento di Fisica dell' Universit\`a di Roma La 
Sapienza'' for kind hospitality.
M.C.  and G.P.  thank the ``Abdus Salam ICTP'' for hospitality during the workshop ``Statistical 
Physics of Frustrated Systems'' 18 August - 7 November 1997.

\end{document}